\documentclass{emulateapj}

\lefthead{van der Wel et al.}
\righthead{$K$-band evolution of early-type galaxies}
\slugcomment{Accepted by Astrophysical Journal Letters}

\begin{document}

\title{The Evolution of Rest-Frame $K$-band Properties of Early-Type Galaxies from $z=1$ to the Present\altaffilmark{1,2,3}}

\author{A.~van~der~Wel\altaffilmark{4}, M.~Franx\altaffilmark{4}, 
P.G.~van~Dokkum\altaffilmark{5},
J.~Huang\altaffilmark{6},
H.-W.~Rix\altaffilmark{7}, and
G.D. Illingworth\altaffilmark{8}}
\altaffiltext{1}{Based on observations collected at the European Southern Observatory, Chile (169.A-0458).}
\altaffiltext{2}{Based on observations with the \textit{Hubble Space Telescope}, 
obtained at the Space Telescope Science Institute, which is operated by AURA, Inc., 
under NASA contract NAS 5-26555}
\altaffiltext{3}{This work is based in part on observations made with the 
\textit{Spitzer Space Telescope}, which is operated by the Jet Propulsion Laboratory,
California Institute of Technology under NASA contract 1407.}
\altaffiltext{4}{Leiden Observatory, P.O.Box 9513, NL-2300 AA, Leiden, The Netherlands}
\altaffiltext{5}{Department of Astronomy, Yale University, P.O. Box 208101, New Haven, CT 06520-8101}
\altaffiltext{6}{Harvard-Smithsonian Center for Astrophysics, 60 Garden Street, Cambridge, MA 02138}
\altaffiltext{7}{Max-Planck-Institut f\"ur Astronomie, K\"onigstuhl 17, D-69117 Heidelberg, Germany}
\altaffiltext{8}{University of California Observatories/Lick Observatory, University of California, Santa Cruz, CA 95064}

\begin{abstract}

We measure the evolution of the rest-frame $K$-band
Fundamental Plane from $z=1$ to the present 
by using IRAC imaging of a sample of early-type galaxies in the 
Chandra Deep Field-South at $z\sim 1$ with accurately 
measured dynamical masses.
We find that $M/L_K$ evolves as 
$\Delta\ln{(M/L_K)}=(-1.18\pm0.10)z$, which is slower than in the $B$-band
($\Delta\ln{(M/L_B)}=(-1.46\pm0.09)z$). In the $B$-band the evolution
has been demonstrated to be strongly mass dependent. In the $K$-band we find a weaker trend:
galaxies more massive than $M=2\times10^{11}M_{\odot}$
evolve as $\Delta\ln{(M/L_K)}=(-1.01\pm0.16)z$; less massive galaxies evolve as
$\Delta\ln{(M/L_K)}=(-1.27\pm0.11)z$.
As expected from stellar population models
the evolution in $M/L_K$ is slower than the evolution in $M/L_B$.
However, when we make a quantitative comparison,  we
find that the single burst Bruzual-Charlot models do not fit the results
well, unless large dust opacities are allowed at $z=1$. 
Models with a flat IMF fit better,
Maraston models with a different treatment of AGB stars fit best.
These results show that the interpretation of rest-frame near-IR photometry is
severely hampered by model uncertainties and therefore that
the determination of galaxy masses from rest-frame near-IR photometry may
be harder than was thought before.
\end{abstract}

\keywords{  cosmology: observations---galaxies: evolution---galaxies: formation }

\section{Introduction}

The formation and evolution of early-type galaxies (hereafter, E/S0s)
has been a major subject of research 
during the past decades. Both theoretical and empirical studies indicate that 
the stellar mass density of the ETG population has increased by at least a
factor of two between $z=1$ and the present day \citep{kauffmanncharlot98,bell04a,faber05}
At the same time, studies at $z\sim 1$ show that massive E/S0s
were already several $Gyr$ old at that epoch \citep{vandokkumstanford03,holden05,treu05b,vanderwel05}. 
It is an intriguing puzzle to explain
these seemingly contradictory results in one picture describing early-type
galaxy formation.

These results hinge on determining galaxy masses, which remains very hard to do at high redshift. 
Typically, mass estimates for high-$z$ galaxies are obtained by comparing their photometric
properties with stellar population models in order to infer a mass-to-light ratio ($M/L$) 
\citep[see, e.g.,][]{bell03,bundy05,shapley05}.
Such SED-fitting methods are intrinsically uncertain, and dynamical mass calibrations are
needed to overcome the large uncertainty 
in the low mass end of the initial mass function (IMF), dark matter content, and, most importantly, 
to verify the validity of stellar population models.
A more quantitative approach is to use the Fundamental Plane
\citep[FP,][]{djorgovskidavis87,dressler87} 
to directly measure the evolution of $M/L$ at rest-frame 
optical wavelengths out to $z=1.3$
\citep{vandokkumfranx96,vandokkum98,kelson00,treu01,vandokkum01,treu02,vandokkumstanford03,vandokkumellis03,
gebhardt03,vanderwel04,wuyts04,holden05,treu05b,vanderwel05}.

Although the measurement of the evolution of $M/L$ itself is model independent, the inferred
age and formation redshift are sensitive
to the choice of the parameters (e.g., IMF and metallicity) of the stellar population model,
and even the choice of the model.
By measuring the evolution of $M/L$ at other wavelengths than the rest-frame $B$-band, the formation redshift
can be established independently, and the range of possible model parameters constrained.
Furthermore, the validity of different models can be verified.
More specifically, different models predict a different evolution of the $M/L$ 
in the near infra-red (NIR) relative to the evolution of the optical $M/L$.
An advantage of measuring the evolution
of the NIR $M/L$ is that the NIR luminosity 
is less affected by the presence of a low mass young stellar population.
Since measuring the evolution of $M/L$ provides a 
luminosity-weighted age estimate, the evolution of the NIR $M/L$ provides a less
biased age estimate.
With the arrival of the near Infra-Red Array Camera \citep[IRAC,][]{fazio04} 
on the Spitzer Space Telescope, 
studying the rest-frame NIR
properties of the high-$z$ E/S0 population has become feasible.
\null
\vbox{
\begin{center}
\leavevmode
\hbox{%
\epsfysize=4.2cm
\epsffile{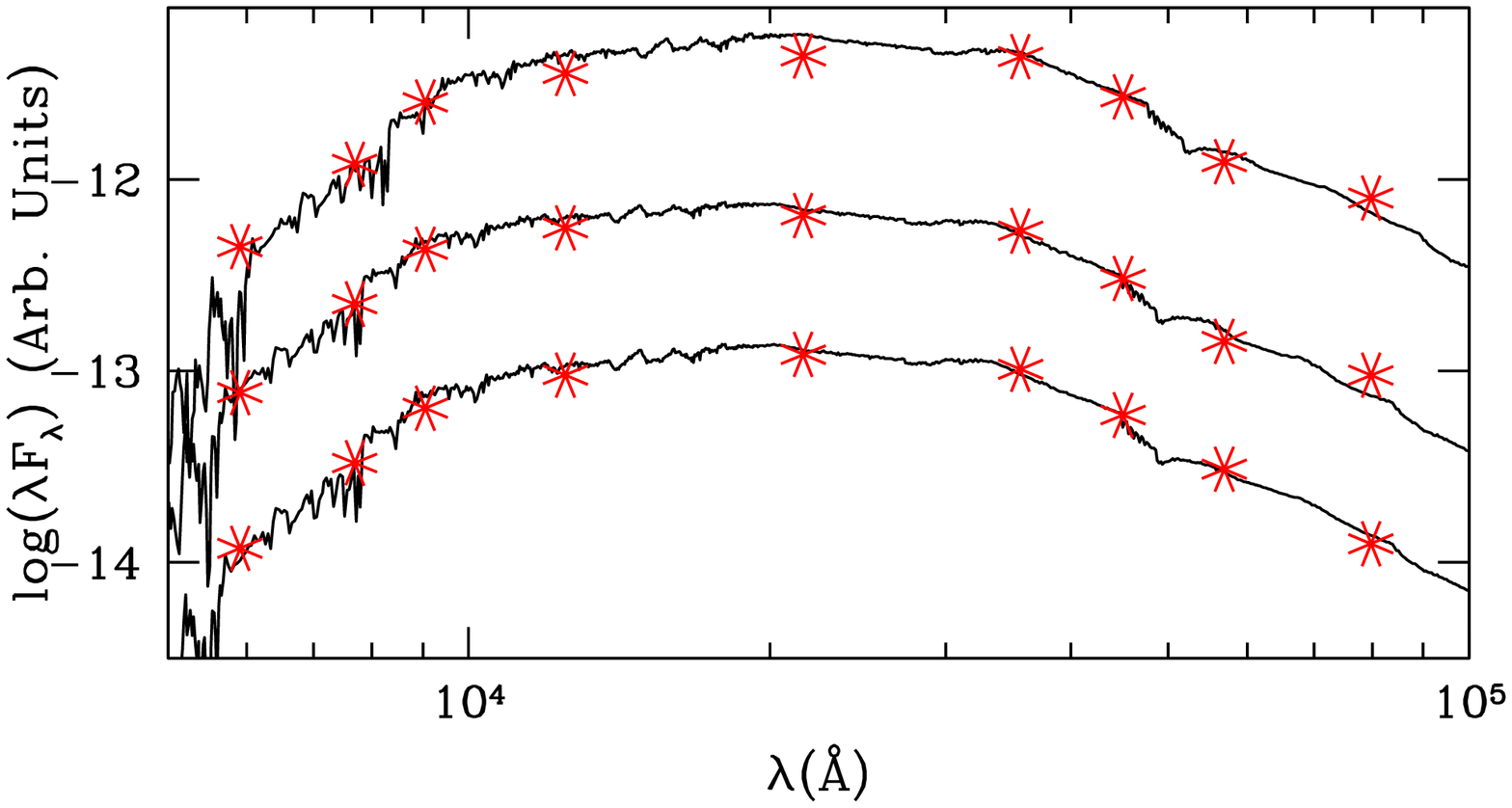}}
\figcaption{\small
SEDs of three $z\sim 1$ E/S0s. The red asterisks are observed fluxes in 
respectively the ACS ($v$, $i$, $z$), ISAAC ($J$, $K$), and IRAC ($3.6$, $4.5$, $5.8$, $8.0$) 
bands.
The lines are model spectra from Bruzual-Charlot
(a $2~Gyr$-old, dust-free SSP with solar metallicity and a Salpeter IMF)
The photometric SEDs, including the IRAC data points, have plausible shapes of stellar populations
of several Gyrs old.
\label{fig:1}}
\end{center}}
In this \textit{Letter} we combine IRAC imaging with our high-$z$ FP study
\citep{vanderwel04, vanderwel05}, such that the evolution of $M/L$ in the rest-frame $K$-band
is measured for the first time. A calibration of galaxy masses derived from NIR luminosities
is thus provided.

\section{IRAC photometry of E/S0s at $z\sim 1$}
\citet{vanderwel05} provide a spectroscopic sample of 29 distant galaxies in the 
CDFS with accurate velocity dispersions. 
Spitzer GTO data are available, providing IRAC imaging in four channels ($3.6\mu$, $4.5\mu$, 
$5.8\mu$, and $8.0\mu$) of a field containing the Chandra Deep Field-South (CDFS).
Using the IRAC data we derive
the rest-frame NIR photometry of those 20 galaxies in the spectroscopic sample with early-type 
morphologies and sufficiently high $S/N$ spectra ($S/N\ge12$ per \AA).
The average redshift of this sample is $z=0.94$ and the average mass is $10^{11} M_{\odot}$. 
12 galaxies in this sample are regarded as the 'primary sample': these objects are all 
at $0.95<z<1.15$ and satisfy all selection criteria applied by \citet{vanderwel05}. 
The other objects, mostly galaxies
at $z\sim 0.7$, are referred to as the 'secondary sample'.
The IRAC data are sufficiently deep (500s) to obtain high $S/N$
photometry of all galaxies in the sample, certainly in the two shortest wavelength channels. 

The IRAC fluxes were measured 
by matching the point spread functions of all available optical/NIR/IRAC images to 
the $8.0\mu$ IRAC image (which has the lowest spatial resolution, $FWHM=2\farcs3$).
$5\farcs0$ diameter aperture fluxes were measured using SExtractor \citep{bertin96} to
derive colors.
Color gradients are ignored, but because the sizes of the $z\sim 1$ galaxies
are much smaller ($r_{eff}\sim 0\farcs4$) than the size of the IRAC PSF
this will only have a very small effect.
The median formal errors in the IRAC fluxes of the galaxies in the sample
are 0.01, 0.01, 0.10 and 0.13 mag for the four channels, respectively.
Additionally, the systematic uncertainty in the IRAC zero-points is 0.03 mag.
We transformed the observed colors to rest-frame
$B-K$ colors using a procedure similar to what is described by \citet{vandokkumfranx96}. 
\null
\vbox{
\begin{center}
\leavevmode
\hbox{%
\epsfysize=4cm
\epsffile{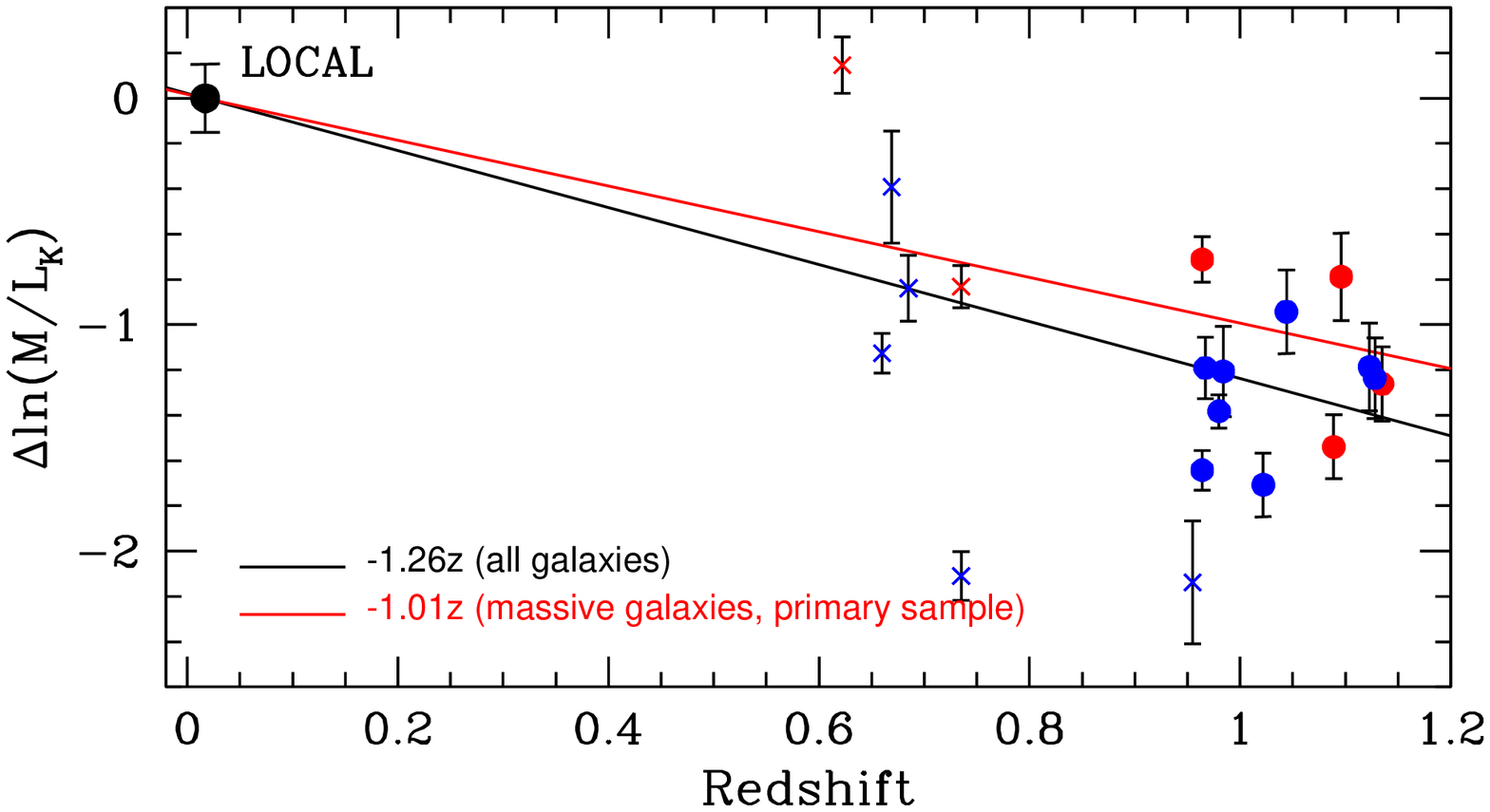}}
\figcaption{\small
$M/L_K$ evolution with redshift of E/S0s in the CDFS.
The solid dots are objects in the primary sample, the crosses are the secondary sample. 
The red symbols are galaxies with high masses
($M>2\times 10^{11} M_\odot$), the blue symbols are less massive galaxies.
The black data point labeled with 'Local' is derived from local cluster 
galaxies, but a small difference between field and cluster
galaxies is taken into account.
The black line indicates the evolution of the entire sample, the red line
indicates the evolution of massive galaxies in the primary sample.
The evolution of $M/L_K$ of massive galaxies is $\sim30\%$ slower
than the evolution of $M/L_B$ of the same sub-sample. 
\label{fig:2}}
\end{center}}
The typical error in the rest-frame $B-K$ color is 0.06 mag, excluding the systematic error of 0.03 mag.
Total luminosities were derived from the structural parameters as obtained from
the ACS $z$-band images and the colors are described above.

From the higher resolution $K$-band image we checked whether the IRAC photometry of our sample
suffers from confusion. Three of the objects in our sample have close neighbors
with such magnitudes that our photometry might be erroneous. However, we do not see
a difference between these contaminated objects and the rest of the sample
in either color, $M/L$, or any other parameter.
In Figure 1 we show the SEDs of three high-redshift E/S0s.
As an illustration we overplot a \citet{bruzual03}
model spectrum for a $2~Gyr$-old stellar population,
demonstrating that the observed SEDs have reasonable shapes. 

We also measured
the colors of a sample of 41 E/S0s in the CDFS 
at $0.6<z<0.8$. 
These were morphologically selected by eye from the ACS imaging.
The redshifts are taken from COMBO-17 \citep{wolf03}.
We use the average $B-K$ color of this sample to complement the $B-K$ evolution
of the galaxies in the FP sample.

\section{Evolution of $M/L_K$}
If galaxies evolve passively, the offset of 
high-$z$ galaxies from the local FP is a measure of the difference
between the $M/L$ of the distant galaxy and the $M/L$ and their local counterparts \citep[see][]{vandokkumfranx96}. 
The rate of evolution is a measure 
of the relative age difference between the local and distant galaxies.

To obtain the $M/L$ evolution in the $K$-band, we compare the NIR properties
of the distant galaxies with the local FP in the $K$-band \citep{pahre98}, i.e.,
we need effective radii ($r_{eff}$) 
and surface brightnesses ($\mu_{eff}$) of the distant galaxies
in the rest-frame $K$-band.
However, the spatial resolution of IRAC is too low to measure 
galaxy sizes at $z\sim 1$. 
Instead we assume that $r_{eff,K}=r_{eff,B}$ 
and determine the surface brightness at this radius.
Although this causes some error in the FP, previous work has shown that the 
combination of $r_{eff}$ and $\mu_{eff}$ which enters the FP is very stable against such
errors \citep{vandokkumfranx96}.
The NIR surface brightness within the optical effective radius is provided
by \citet{pahre98} for a local sample of cluster galaxies. 
For the distant sample we compute $\mu_{eff,K}$ from 
$\mu_{eff,B}$ and $(B-K)_{eff}$, where $(B-K)_{eff}$ is the color inside the effective
radius. It is calculated from the total $B-K$ color by 
correcting for the negative color gradient measured by \citet{peletier90}.
The difference between $B-K$ and $(B-K)_{eff}$ is -0.07 mag.

The local $K$-band FP is based on cluster galaxies, but our sample 
consists of field galaxies. 
\begin{figure*}[t]
\begin{center}
\leavevmode
\hbox{%
\epsfxsize=6.9cm
\epsffile{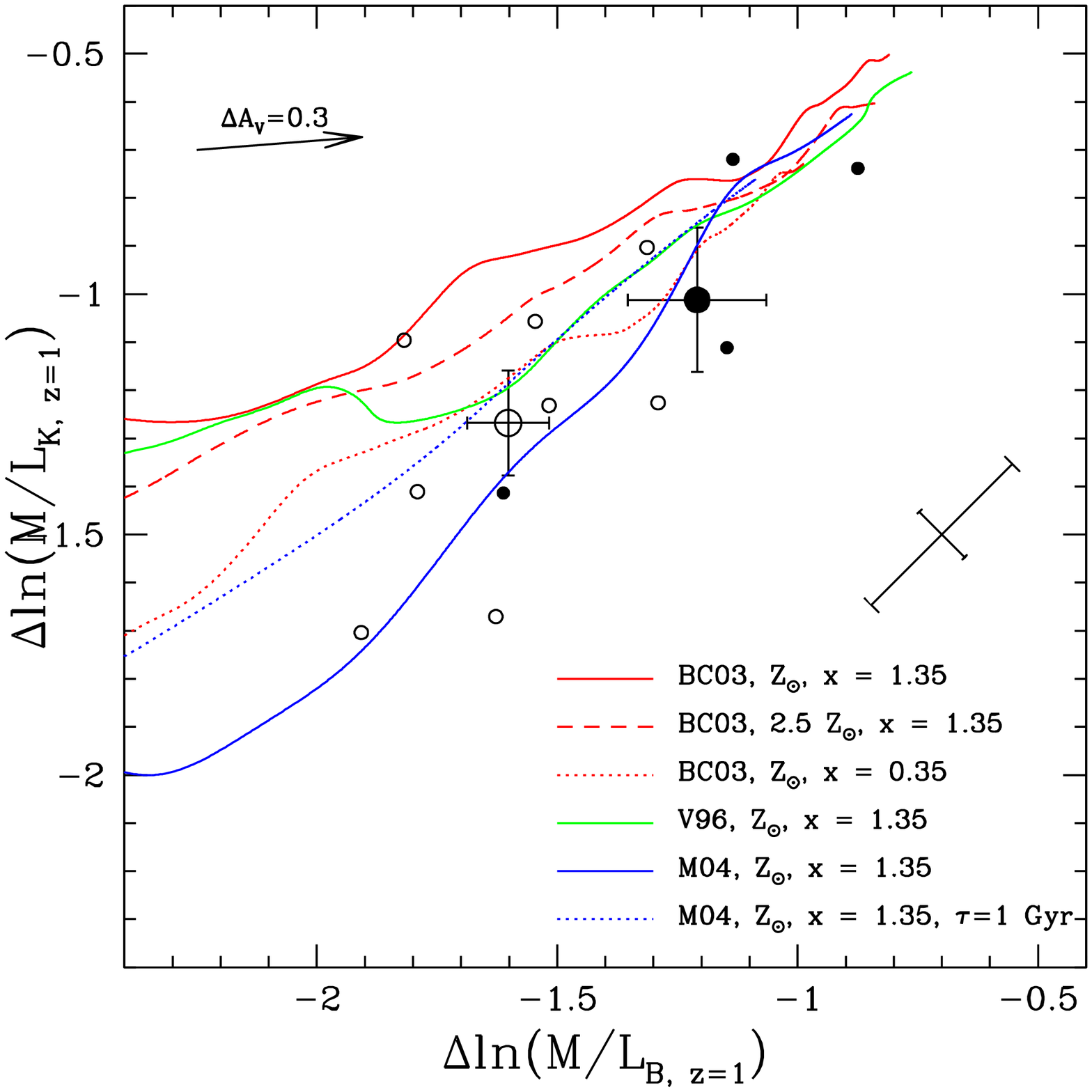}
\epsfxsize=6.9cm
\epsffile{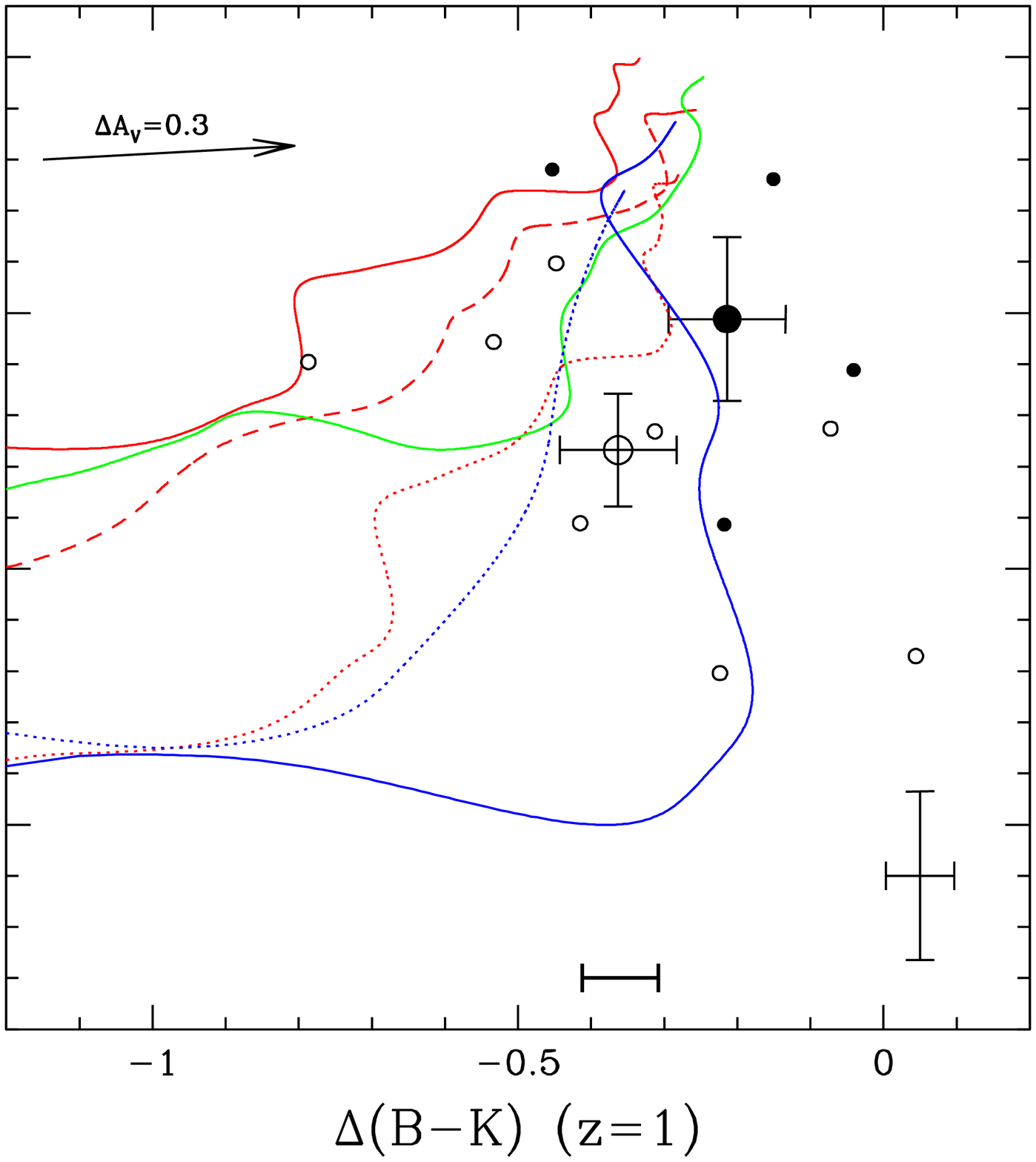}}
\figcaption{
\small
Comparison of the $M/L$ evolution in the rest-frame $B$- and $K$-bands
and the $B-K$ colors of our $z\sim 1$ E/S0 sample.
(Left): the evolution of $M/L_B$ versus the evolution of $M/L_K$, normalized at $z=1$.
The large filled symbol with error bars is the average of the four galaxies
in the primary sample with masses larger than $M=2\times 10^{11} M_\odot$. The large
open symbol with error bars represents the eight less massive galaxies in the primary sample.
The smaller symbols are the individual galaxies in the primary sample. 
The diagonal error bar at the right shows the typical errors
in  $\Delta \ln{M/L_B}$ and  
$\Delta \ln{M/L_K}$ for the individual data points. 
The shorter side of the error bar represents the uncertainty
in $B-K$. The vector indicates the effect of dust.
(Right): the evolution of $B-K$ versus the evolution of $M/L_K$, normalized at $z=1$.
The thick horizontal error bar at the bottom indicates the average $B-K$ evolution extrapolated
to $z=1$ of a sample of 41 E/S0s at $0.6<z<0.8$.
The model tracks indicate the expected evolution between $z=1$ and the 
present, varying
the formation redshift along the tracks.
The Bruzual-Charlot models with a Salpeter IMF fit badly to the data.
A Bruzual-Charlot model with a flat IMF provides a better fit.
The Maraston models with a Salpeter IMF provides the best fit.
These models have a different treatment of the AGB stars.
\label{fig:3}}
\end{center}
\end{figure*}
\citet{faber89} have shown that field galaxies have $\sim 5\%$
lower $M/L_B$ than cluster galaxies. Assuming that this difference is caused by either
a difference in age or metallicity, this translates into a 2\% difference in $M/L_K$, which
follows from stellar population models from \citet{bruzual03}.
We take this difference into account when comparing the FP at high and low redshifts.

We show the evolution of $M/L_K$ with redshift in Figure 2. It is apparent
that $M/L_K$ evolves significantly from $z\sim 1$ to the present. The evolution of the
sample is $\Delta \ln{(M/L_K)}=(-1.26\pm0.15)z$,
obtained from a least squares linear fit.
The primary sample alone evolves at a similar rate: 
$\Delta \ln{(M/L_K)}=(-1.18\pm0.10)z$.
Comparing these numbers to the evolution of $M/L_B$
($\Delta \ln{(M/L_B)}=(-1.52\pm0.16)z$ and $(-1.46\pm0.09)z$, respectively),
we see that the evolution in $M/L_K$ is somewhat slower than 
the evolution in $M/L_B$.
The scatter in $\Delta \ln{(M/L_K)}/z$ is $0.32$, which is not much
smaller than the scatter in the $B$-band ($0.37$).
This is much larger than can be accounted for by measurement errors, 
and is most likely caused by an age spread
of the stellar populations of the galaxies, as has been demonstrated before 
\citep{vanderwel05}.
In the $B$-band the measured evolution is strongly mass-dependent
\citep{vanderwel04,treu05b,vanderwel05}, mostly due to selection
effects, but partially due to intrinsic differences between high- and low-mass 
galaxies \citep{vanderwel05}.
In the $K$-band the difference is much less pronounced, and only marginally significant. 
The galaxies in the primary
sample with masses higher than $M=2\times 10^{11} M_{\odot}$ evolve as 
$\Delta \ln{(M/L_K)}=(-1.01\pm0.16)z$; galaxies less massive than that
as $\Delta \ln{(M/L_K)}=(-1.27\pm0.11)z$. 
In the $B$-band these numbers are,
respectively, $(-1.20\pm0.14)z$ and $(-1.60\pm0.09)z$.
These numbers are consistent with the model prediction 
that $M/L_K$ is less sensitive than $M/L_B$ 
to the age of a stellar population and recent star formation.

Concluding, we find that the evolution of $M/L_K$ from $z=1$ to the present
is $\sim30\%$ slower than the evolution of $M/L_B$. 
We note that this direct measurement deviates from previous determinations
of the evolution of $M/L_K$ that were based on extrapolating observed SEDs
(including the $K$-band as the longest wavelength data)
to the rest-frame $K$-band at $z\sim 1$ using stellar population models.
Such studies found that $M/L_K$ evolves at least twice as slow as $M/L_B$
\citep[see, e.g.,][]{drory04}.
This demonstrates that extrapolating 
observed $K$-band photometry of high-$z$ galaxies to rest-frame $K$-band $M/L$ 
is hazardous and strongly model dependent.

\section{Discussion}
Our results allow us to compare the evolution of $M/L_B$ and $M/L_K$ directly with predictions
from stellar population models.
We compare our results with the predictions of three models:
\citet{bruzual03}, \citet{vazdekis96}, and \citet{maraston04} (hereafter, BC03, V96, and M04, respectively). 
A critical aspect of the models
is the method that is used to implement late stellar evolutionary phases. 
BC03 and V96 compute isochrones 
up to the early AGB phase, and include an empirical prescription
for the thermally pulsating- (TP-) AGB phase. On the contrary, M04 adopt the
'fuel consumption' approach, which allows implementation
of short-duration but very luminous evolutionary stages, such as the TP-AGB 
phases, in an analytical and numerically stable way.

In Figure 3 we compare the evolution from $z=1$ to the present of $M/L_B$ and $M/L_K$,
and the related change in the rest-frame $B-K$ color.
The BC03 model with solar metallicity and a Salpeter IMF does not fit
the data: the predicted evolution of $M/L_K$ against $M/L_B$ is significantly slower 
than observed.
Before investigating different models and model parameters, we 
consider several possible explanations 
for an apparently fast evolution of $M/L_K$ with respect to $M/L_B$
that are unrelated to stellar populations.
First, a difference in dust content between 
local E/S0s and $z\sim 1$ E/S0s could lead to redder
colors at  $z=1$  than expected from dust-free models.
However, the amount of reddening required to match the BC03 model
with solar metallicity and a Salpeter IMF
is considerably large for half of the galaxies in our sample ($A_V> 0.5$),
and several even require $A_V> 1$.
Although the optical spectra show no indication of such high opacities,
observations at longer wavelengths with Spitzer can constrain the
presence of dust-enshrouded populations which might contribute in the K-band.
One might expect to observe irregular morphologies or irregular intensity
profiles if the absorption were so high. We note that the galaxies
do not show such effects at the ACS resolution.
Second, the red colors might be produced by AGN in the $z\sim 1$ sample
that are obscured in the optical, but not entirely in the NIR.

Next, we test whether changing parameters within the single-burst
BC03 model can improve
the consistency between the model and the data.
Increasing the metallicity does not change the predictions by much. 
Changing the star formation history (SFH) does not lead to better fits to the evolution of 
$M/L_B$ and $M/L_K$, unless models with constant star formation rates or
exponentially declining star formation rates with e-folding time scales 
of several $Gyr$ are adopted. These models, however, have very blue colors
that do not match the $B-K$ colors of the galaxies in our sample, nor
the colors of the local population.
Flattening the slope of the IMF (from 1.35 to 0.35)
does shift the model curve toward our data points, leading to an
acceptable  fit.

Finally, we
compare our results to other models.
As can be seen in Figure 3, the V96 and M04 models with a Salpeter IMF and solar
metallicity provide better fits than BC03. 
Note that the differences between the 
models with identical parameters are very large;
e.g., BC03 predict a correlation between color and $\Delta \ln{(M/L)}$, 
whereas M04 does not.
We note that the spread of the data points representing the individual galaxies is large,
but that there is no clear correlation between $\Delta \ln{(M/L_K)}$ and $\Delta (B-K)$.
Even assuming a flat IMF, the BC03 model cannot reproduce the data points at the bottom right 
part of Figure 3b.
The M04 model does fit to those data points. On the other hand, data points at the upper left
of Figure 3b are not fit well by the M04 model.
However, all but one of the these data points can be fit with a model
with an exponentially declining star formation rate with $\tau=0-1~\rm{Gyr}$.

We can use the simple single burst models to estimate the average, luminosity-weighted
formation redshift of the stellar populations of the galaxies.
The formation redshift as estimated from the evolution of $M/L_B$ 
is $1.6\leq z_{form}\leq 1.9$
and $1.5\leq z_{form}\leq 1.9$, using the models from BC03 and M04, respectively, 
with a Salpeter IMF and solar metallicity.
According to the same models, the evolution of $M/L_K$ suggests that 
$1.2\leq z_{form}\leq 1.5$ and $1.6\leq z_{form}\leq 2.0$, respectively.
The BC03 model with a flat IMF provides a better fit to our results than the BC03
model with a Salpeter IMF and produces $1.7\leq z_{form}\leq 2.7$, both from the
evolution in the $M/L_B$ and the evolution in $M/L_K$.
As can be seen, the constraints on $z_{form}$ are not improved by measuring the evolution
of $M/L_K$. This is entirely due to the rather large model uncertainty in the rest-frame 
$K$-band evolution.
Obviously, it is of the greatest relevance to improve our knowledge
of both the AGB phase and the dust content of high redshift elliptical
galaxies, as dust may mimic some of the results we see.

Finally, we note that E/S0s are thought to be relatively simple systems, with simple SFHs. 
It is likely that the model uncertainties are even higher for galaxies with a more complex SFH.
Irrespective of the causes of the discrepancies between the models and the observations,
the results presented here suggest that the rest-frame $K$-band photometry
of such galaxies may be difficult to interpret.
Further studies of the role of dust and AGN would be valuable.
If they do play a role in the photometric bands analyzed here,
proper modeling of the colors is more complex than thought before.
In the next paper, we will explore full SED fitting, the uncertainties in determining stellar masses,
and distinguish between the applicability of different stellar population models.

The authors would like to thank Stijn Wuyts and Ivo Labb\'e for discussing photometry on IRAC data
and the Leidsch Kerkhoven-Bosscha Fonds for financial support.
PGvD acknowledges support from NASA LTSA grant NNG04GE12G.

\end{document}